\documentclass{csmagclass2}														

\begin{document}		
\headings																															 %

\title{{Area-Law Study of Quantum Spin System\\ on Hyperbolic Lattice Geometries}}
\author[]{{A. GENDIAR\thanks{Corresponding author: andrej.gendiar@savba.sk}}}
\affil[]{{Institute of Physics, Slovak Academy of Sciences, D\'{u}bravsk\'{a} cesta 9, 845 11 Bratislava, Slovakia}}

\maketitle

\begin{Abs}
Magnetic properties of the transverse-field Ising model on curved (hyperbolic) lattices are studied by a tensor product variational formulation that we have generalized for this purpose. First we identify the quantum phase transition for each hyperbolic lattice by calculating the magnetization. We study the entanglement entropy at the phase transition in order to analyze the correlations of various subsystems located at the center with the rest of the lattice. We confirm that the entanglement entropy satisfies the area law at the phase transition for fixed coordination number, i.e., it scales linearly with increasing size of the subsystems. On the other hand, the entanglement entropy decreases as power-law with respect to the increasing coordination number.
\end{Abs}
\section{Introduction}
The tensor-product states have been intensively studied in various strongly correlated systems, focusing mainly on their ground-state properties~\cite{Orus,Fra}. They were developed for treating two-dimensional quantum systems and are of intensive interest in the recent decade. We consider a simple spin model in order to study magnetic properties on non-Euclidean lattices, which describe negatively curved two-dimensional surfaces. Such spin lattices are generally known as the hyperbolic lattices and can also be understood as generalizations of the Bethe lattices. These lattices can form so-called {\it spin networks}, which are meant to describe anti-de Sitter (i.e., hyperbolic) spaces used in theory of quantum gravity~\cite{Mal}. By analyzing the magnetization and the entanglement entropy on the hyperbolic lattices, we connect the solid-state viewpoint with the correspondence between the Anti-de Sitter (AdS) and the conformal field theory (CFT), which is specified in quantum gravity.

Since the quantum spin systems on arbitrary hyperbolic lattices are not analytically solvable, we treat them numerically by a generalized tensor-network algorithm, Tensor Product Variational Formulation (TPVF). We have successfully applied the TPVF method to the Heisenberg, XY, and Ising models~\cite{Miso}. It is worth mentioning that the classical spin analogs have also been intensively studied on the hyperbolic surfaces~\cite{3q,pq}, as they exhibit many interesting features, which they have in common with quantum spin systems. As the simplest example, we consider the quantum Ising model on infinitely large hyperbolic lattices in order to observe phase transitions after the spontaneous symmetry-breaking occurs. The hyperbolic lattices are constructed by regular tessellation of identical polygons with uniform coordination numbers. They form negatively curved surfaces with constant Gaussian curvatures and infinite Hausdorff dimension.

\section{Model and Method}
\begin{figure}[tb]
\begin{center}
\phantom{.}\hfill
\includegraphics[width=3.0cm]{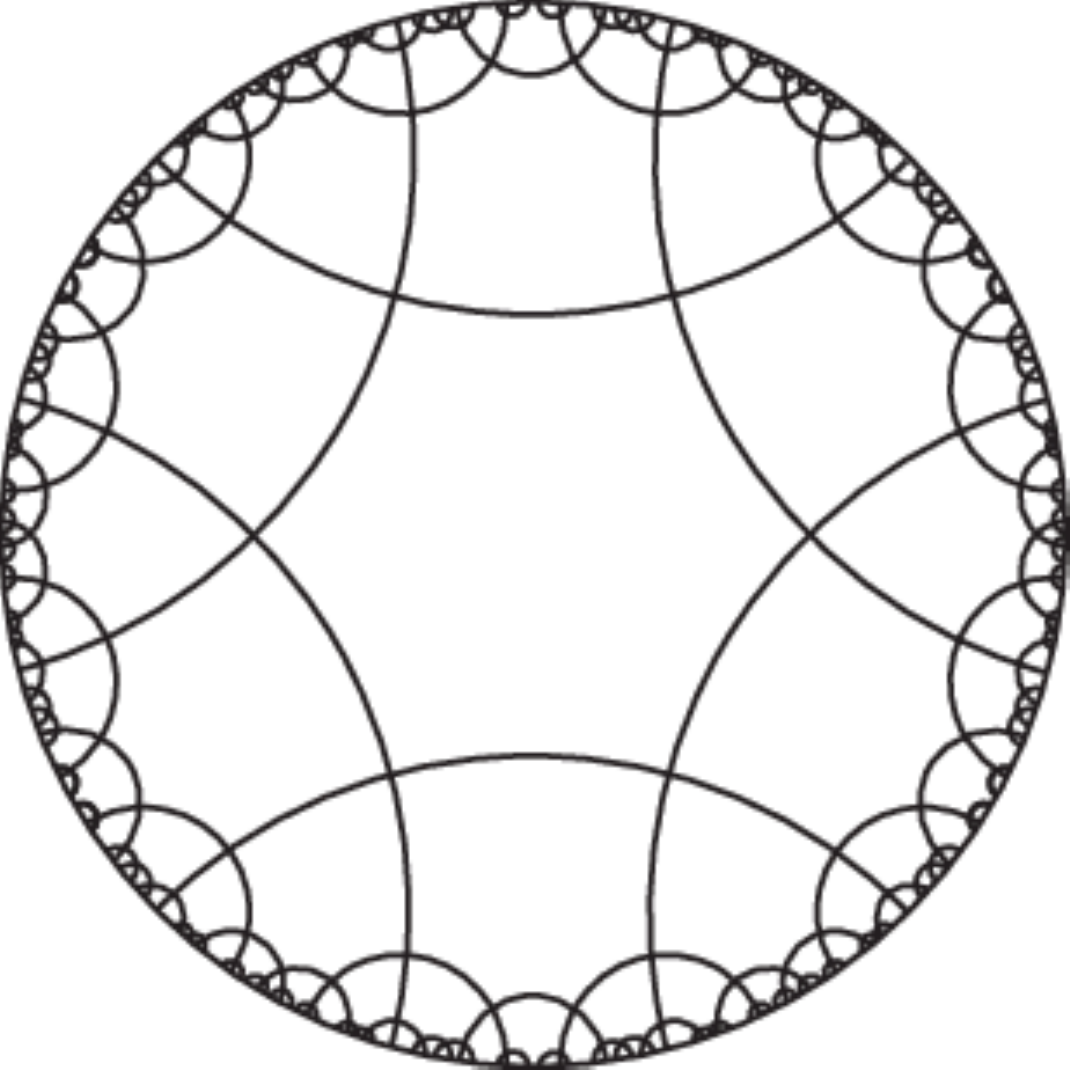}$(6,4)$
\hfill
\includegraphics[width=3.0cm]{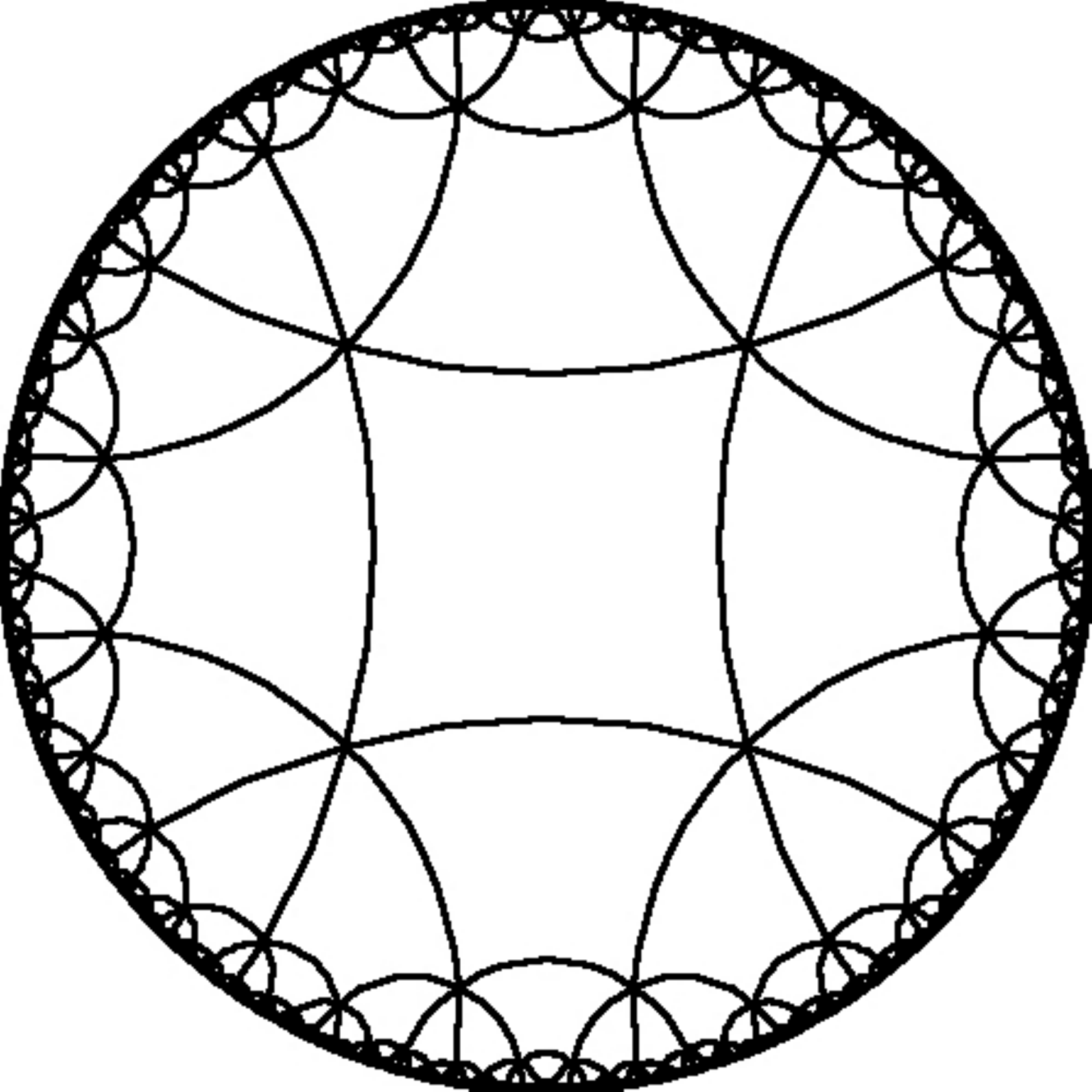}$(4,6)$
\hfill\phantom{.}
\end{center}
\caption{Example of the hyperbolic lattice geometries $(6,4)$ and $(4,6)$. The lattices are shown in the Poincar\'{e} disk representation, which is a mapping of the identical polygons onto the disc.}
\label{fig1}
\end{figure}

The main aim of this study is to revisit critical properties of the quantum spin systems with respect to underlying lattice surfaces. Let us consider the transverse-field Ising model
\begin{equation}
	{\cal H}_{(p,q)}=-J\sum\limits_{\{i,j\}_{(p,q)}} S^z_i S^z_{j}
	-h^x_{~} \sum\limits_{\{i\}_{(p,q)}} S^x_i \, ,
	\label{Hm}
\end{equation}
where the Pauli matrices $S^z_i$ and $S^x_i$ are located on a site $i$ of particular hyperbolic lattice $(p,q)$, which is characterized later. We consider ferromagnetic coupling $J>0$ acting between nearest neighbors $\{i,j\}$, and constant transversal magnetic fields $h^x_{~} \geq 0$. The summation runs over all spin sites $\{i\}$ on the negatively curved surfaces. They represent a class of the uniform hyperbolic lattices~\cite{pq}. These hyperbolic lattices are usually classified by a pair of two positive integers $(p,q)$ and we consider the case when $p,q\geq4$. The first integer describes a regular polygon with $p$ sides (e.g., $p=4,5,6,\dots$ correspond to the square, pentagon, hexagon, etc.) while the second integer is the coordination number $q$ which is kept uniform on the entire hyperbolic lattice. Hence, $(p\geq4,q\geq4)$ describes an infinite set of the hyperbolic lattices with the only exception: the $(4,4)$ geometry refers to the flat square lattice. For instance, we depict two hyperbolic lattices in Fig.~\ref{fig1} in the so-called Poincar\'{e} disk representation~\cite{Poinc}. The $(6,4)$ lattice is created by the regular tessellation of identical hexagons, $p=6$, with the uniform coordination number $q=4$, whereas the dual $(4,6)$ lattice is made by tiling the regular squares, $p=4$, with $q=6$.

The infinite size of the hyperbolic lattice geometries $(p,q)$ is necessary for studying the phase transitions. The ferromagnetic Ising model exhibits a single second-order (continuous) phase transition, which separates the ferromagnetically ordered phase from the disordered phase at a certain quantum phase-transition field $h^x_{\rm pt}$. For instance, the Ising model has a phase transition at $h^x_{\rm pt}=3.0439$ on the square $(4,4)$ lattice~\cite{HOTRG}. To find the phase transition, we calculate the spontaneous magnetization $\langle S^z_{~}\rangle = \langle \Psi_0 \vert S^z_{~} \vert \Psi_0 \rangle$, as the relevant order parameter. Here, the wave function $\vert\Psi_0\rangle$ is the ground state corresponding to the lowest energy $E_0$ of the Hamiltonian ${\cal H}_{(p,q)}$, we have specified in Eq.~(\ref{Hm}). The transverse magnetic field, at which the non-zero magnetization $\langle S^z_{~}\rangle$ tends to zero value, specifies the phase transition we denote by $h^x_{\rm pt}$ in the following.

There is another important quantity, which is also useful in precise detection of the phase transition. It is the entanglement entropy $S_{(p,q)}=-{\rm Tr}\left(\rho\log_2\rho\right)$, which gets maximized at phase transitions. To calculate the entanglement entropy $S_{(p,q)}$, we need to obtain a reduced density matrix $\rho={\rm Tr}^{\prime} \vert\Psi_0\rangle \langle\Psi_0\vert$ by partial tracing out the environment of the entire lattice system described by $\vert\Psi_0\rangle$. The entanglement entropy $S_{(p,q)}$ can represent a function which quantifies the amount of quantum correlations of a certain subsystem coupled to the rest of the system (the reservoir). We consider such subsystems, which are formed by regular polygons of $p$ sides and are located in the center of the hyperbolic lattice (deeply in the bulk). The reservoir is the remaining part of the hyperbolic lattice. When evaluating the entanglement entropy, all degrees of the freedom belonging to the reservoir are traced out. To accomplish these calculations, we employ the TPVF method. The method has been found reliable, because it approximates the ground state $\vert\Psi_0\rangle$ correctly by means of the tensor product~\cite{Miso}. We use the method to obtain the spontaneous magnetization $\langle S^z_{~}\rangle$ and the entanglement entropy $S_{(p,q)}$.
\section{Results}
\begin{figure}[tb]
\includegraphics[width=0.48\textwidth]{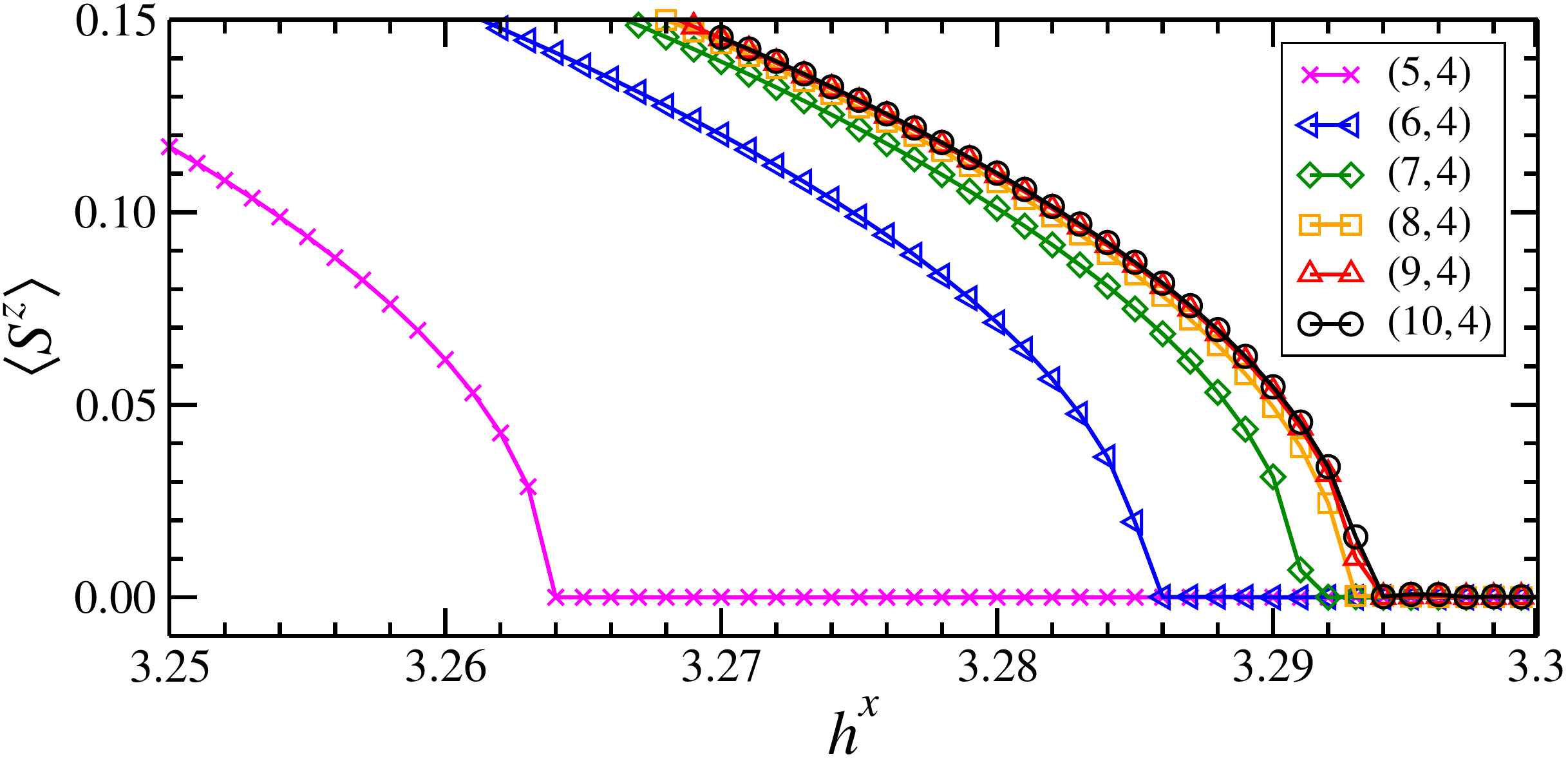}
\hfill
\includegraphics[width=0.48\textwidth]{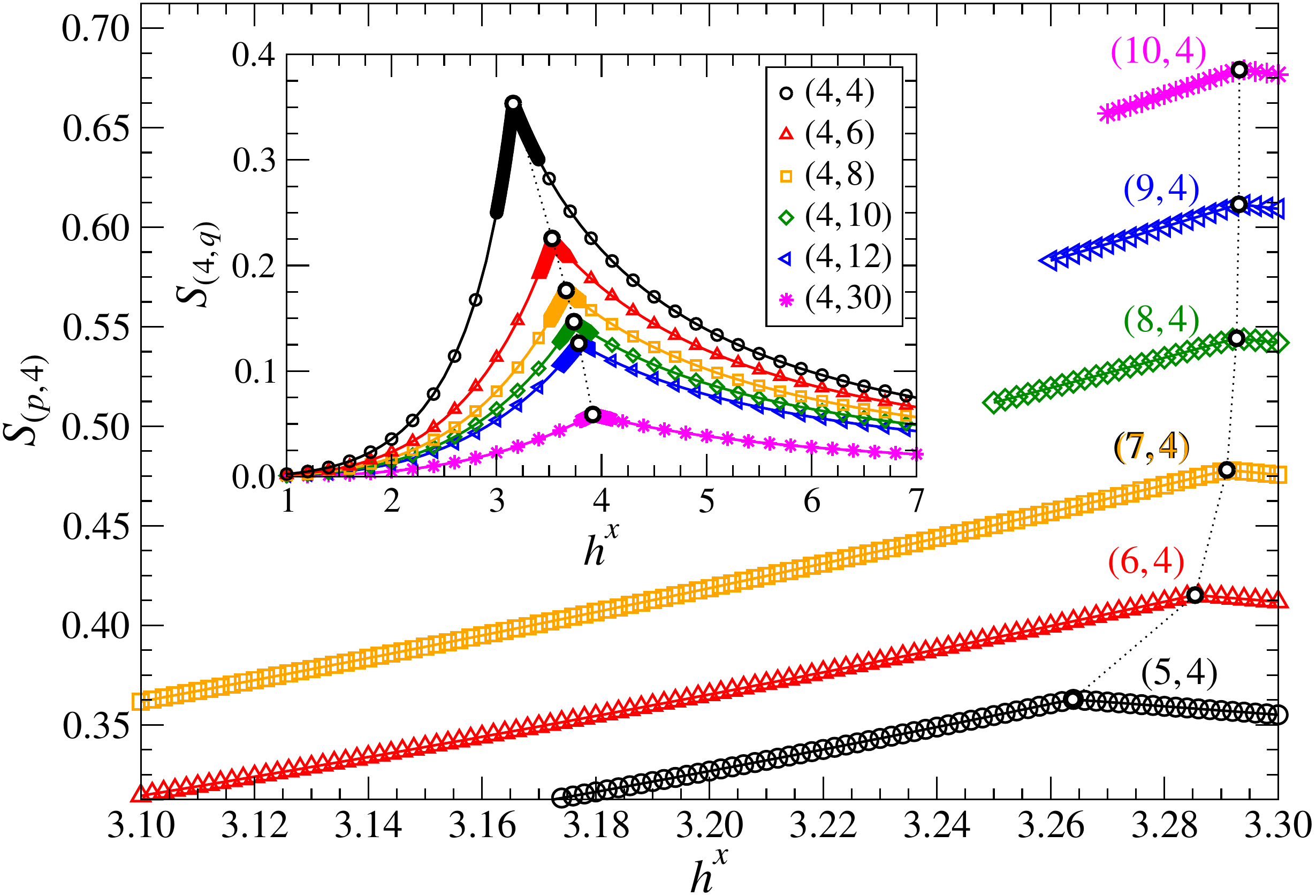}
\caption{Left: Magnetization $\langle S^z_{~} \rangle$ as the order parameter with respect to the transversal magnetic field $h^x_{~}$ on the hyperbolic lattices $(p,4)$. Right: Entanglement entropy versus the transversal field $h^x$. The maxima of $S_{(p,4)}$ at phase-transition field $h^x_{\rm pt}$ (the open circles) increase with $p$ at $q=4$, whereas the inset shows that the maxima of $S_{(4,q)}$ decrease with $q$ at fixed $p=4$.}
\label{fig2}
\end{figure}
The quantum phase-transition field $h^x_{\rm pt}$ can be numerically obtained for sufficiently large lattices after all thermodynamic functions completely converge. This is carried out iteratively within TPVF, which breaks the spin symmetry resulting in a non-zero spontaneous magnetization. The method produces a smooth dependence of the magnetization on the transversal field $h^x_{~}$ with the well-defined phase-transition field $h^x_{\rm pt}$. The ordered phase with the non-zero magnetization $\langle S^z_{~}\rangle>0$ for $h^x_{~}<h^x_{\rm pt}$ is separated from the disordered phase with $\langle S^z_{~}\rangle=0$ for $h^x_{~}\geq h^x_{\rm pt}$. Figure~\ref{fig2} (left) shows the magnetic-field dependence of the magnetization for hyperbolic lattices $(p,4)$ made of the regular polygons $5\leq p\leq 10$ with the coordination number $q=4$. As $p$ increases, the $h^x_{\rm pt}$ rapidly converges to the value $h^x_{\rm pt}=3.2922$, which corresponds to the Bethe lattice with $q=4$.

The identical phase-transition fields $h^x_{\rm pt}$ can be reproduced if observing the maxima of the entanglement entropy $S_{(p,q)}$ plotted in Fig.~\ref{fig2} (right) for the same set of the hyperbolic lattices with $5 \leq p \leq 10$ at $q=4$. The maxima of $S_{(p,4)}$ corresponds to identical phase transitions $h^x_{\rm pt}$. They are marked by the open circles, and $h^x_{\rm pt}$ rapidly saturates as we have seen for $\langle S^z_{~} \rangle$. The scaling of the maxima in $S_{(p,4)}$ is discussed below.

On the contrary, the entanglement entropy shows a substantially different dependence if $q$ varies and $p$ is fixed. In particular, the polygons describe the squares ($p=4$) while the coordination number $q$ grows, as shown in the inset of Fig.~\ref{fig2} (right). The maxima of the entanglement entropy $S_{(4,q)}$ are again associated with the phase-transition fields $h^x_{\rm pt}$ (marked by the open circles). Here, however, the maxima of $S_{(4,q)}$ decrease as $q$ increases. If taking the limit $q\to\infty$ for $p=4$, we obtain $h^x_{\rm pt}=4$ after extrapolation (not shown). The suppression of the entropy $S_{(4,q)}$, while increasing the coordination number $q$, suggests that the correlations on the hyperbolic lattices with large $q$ can significantly weaken down to zero even at phase transition.

\begin{figure}[tb]
\centerline{\includegraphics[width=0.47\textwidth]{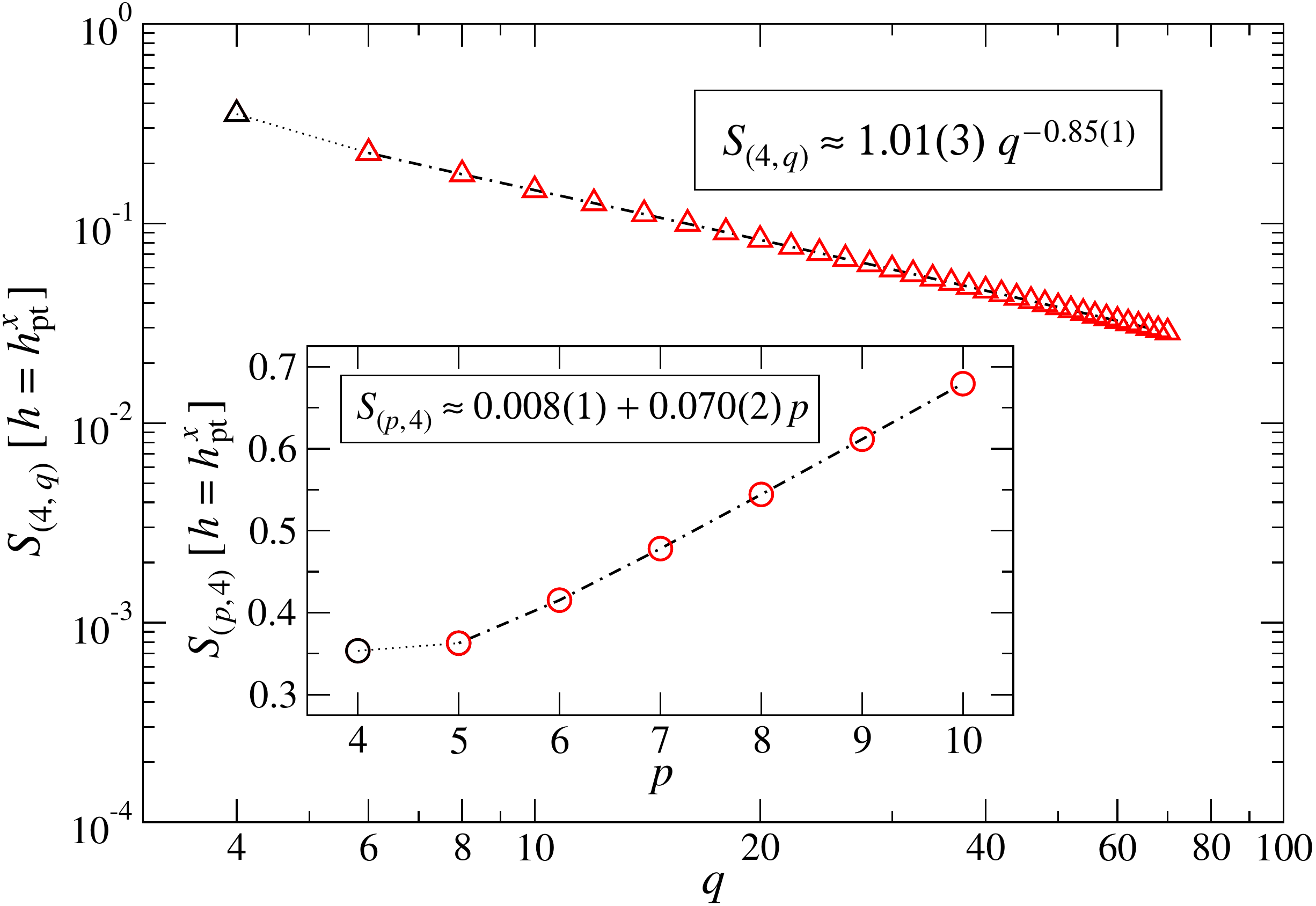}}
\caption{The entanglement entropy $S_{(4,q)}$ with respect to $4\leq q\leq70$ in the log-log scale follows the power-law behavior $S_{(4,q)}\approx1.01(3)/q^{0.85(1)}$. Inset: Linear dependence of $S_{(p,4)}\propto p$.}
\label{fig3}
\end{figure}
Thus, the $(p,q)$ dependence of the entanglement entropy $S_{(p,q)}$ at the phase transitions $h^x_{\rm pt}$ suggests the two above-mentioned scenarios. The hyperbolic lattices $(p,q)$ can be analytically described by the (negative) Gaussian curvature~\cite{pq}, which decreases if both $p$ and $q$ increase (we remark here that the $q$-dependence affects the curvature stronger than the $p$-dependence).

Figure~\ref{fig3} shows the linear decrease of the entanglement entropy with respect to $q=4,5,\dots,70$ in the log-log scale, i.e., the power-law dependence of the entropy obtained by the least-square fitting yields $S_{(4,q)}\approx q^{-0.85}$. (The square lattice $(4,4)$ data were eliminated from the fitting.) If the coordination number $q=4$, the increase of the polygon sides $p=4,5,6,\dots$ leads to linearity of the entanglement entropy $S_{(p,4)}\approx p$, see the inset of Fig.~\ref{fig3}. This linearity supports the validity of the area law for the hyperbolic lattices. The area law states that the entanglement entropy scales with the surface size of the subsystem (not as the volume of the subsystem)~\cite{Jens} and has not been studied for the non-Euclidean systems yet.

\section{Conclusions}
We analyzed the entanglement-entropy scaling for the transverse-field Ising model at its phase transition with respect to small subsystem sizes parameterized by $p$ and $q$. We conjecture that the entaglement entropy decreases algebraically with respect to $q$ (for fixed $p=4$), whereas the area-law scaling, $S \sim p$, is preserved only for fixed coodination number (we used $q=4$), which can be considered as being one of the building blocks within the AdS-CFT correspondence.

\section{Acknowledgement}
I would like to express my thanks to Michal Dani\v{s}ka. The projects APVV-16-0186 (EXSES), APVV-18-0518 (OPTIQUTE), VEGA 2/0123/19, and JTF QISS are greatly acknowledged.
%


\end{document}